\documentclass[twocolumn,showpacs,floatfix,eqsecnum,prb]{revtex4}

\usepackage[dvips]{epsfig}

\usepackage{float}

\begin{document}

\title{On the single-electron theory of quantum spin Hall effect in two dimensional topological insulators}
\author{Yi-Dong Wu}
\affiliation{Department of Applied Physics, School of Science, Yanshan University, Qinhuangdao, Hebei, 066004, China}
\email{wuyidong@ysu.edu.cn}

\begin{abstract}

Recently we wrote a paper on the theory of the quantum spin Hall effect(QSHE) in two dimensional(2D) topological insulators(TIs)\cite{Wu2011a} which have been considered as ``do not add much new insight to the exhaustively studied topic of TI within a single-electron picture'' by the referees. In this paper we review the papers on the mechanism of the QSHE which have significant influence on understanding of the subject. By illustrating the failures of the previous works we show our paper do contribute a different point of view to this topic, which we believe is not only a new but also the correct way to approach the problem at the single-electron level.
\end{abstract}

\maketitle
The main idea of our recent paper is that the 2D topological insulator can be viewed as two time-reversal related Chern insulators with opposite odd Chern numbers. The two Chern insulators can continuously pump electrons to opposite directions, thus produce QSHE. Contrary to the $Z_2$ picture, our conclusion is the isolated 2D TI won't return to its origin state after two cycles and QSHE can be observed in isolated 2D TIs. We proposed an experiment two confirm our assertion. In this paper we first demonstrate why the $Z_2$ picture and some other pictures fail, then we elaborate our point of view in some detail.\\
Admittedly, the QSHE in 2D TI have been extensively studied, we don't believe the researches are exhaustive even at the single-electron level and there are many technical flaws in the existing theoretical works. We totally agree with the referee about ``the idea that a $Z_2$ TI can be regarded as pair of parallel $n=1$ quantum Hall states, with opposite
Chern number for opposite spins, has been extensively discussed in earlier literature''.\cite{Kane2005a,Bernevig2006} However, this idea is only popular at the early age of the field when we considered the simple situation where one component of the spins is conserve. In this simple case the eigenstates of the Hamiltonian have definite spin and the occupied bands with opposite spin naturally form a time-reversal related pair. When the spin non-conserving terms are introduced the Chern number picture become out of vogue\cite{Kane2005b}. One statement capture the change of the attitude towards the Chern number come from an review article which we quote here ``While $n_\downarrow$, $n_\uparrow$ lose their meaning when $S_z$ nonconserving terms (which are inevitably present) are added, $\nu$ retains its identity''\cite{Hasan2010}. In this paper we will show $n_\downarrow$, $n_\uparrow$ don't simply lose meaning, instead, they evolve to a new kind of Chern numbers which determine the behavior of the 2D TIs when $\hat{s_z}$ is no longer conserve. \\
Several authors have explain the QSHE in 2D TIs by studying the cycles or electron pumps of the 1D systems. Here we first show those explanations don't work.\\
\section{The difficulties of the pumps of 1D system}
Topological band theory and topological field theory are two most popular theories of the TIs on the single-particle level. Both of theories have tried to explain the QSHE in 2D TIs by studying the adiabatical pumping cycles of the 1D system in their representative works\cite{Fu2006,Qi2008}. Now we show 1D systems don't work as they have described.
\subsection{An obvious mistake }
In Ref\cite{Qi2008} a two-band tight-binding model is studied to review the physics of the time reversal-breaking(TRB) topological insulator and illustrate the procedure of dimension reduction. They proposed an electron pump and calculate the charge flow in one cycle. When they tried to understand the quantum Hall response in the 1D picture they made an obvious and serous mistake. It's claimed that ``the charge is always pumped to the left for the half-filled system'' and the edge states evolution for $k_y=0\rightarrow 2\pi$ is illustrated in the Fig. 1(c) in Ref \cite{Qi2008}. Unfortunately both the conclusion and the the figure are wrong. The reason is that the edge states don't behave as drawn in the figure. First, the edge states don't always exist for all $k_y=0\rightarrow 2\pi$ as indicated in the Fig 1(b) in Ref \cite{Qi2008}. In fact only the two cross lines in the rectangle in FIG. 1(a) represent edge states. In FIG.1(b) and (c) we draw the average x position and the position uncertainty in the x direction of the eigenstates. It's easy to see only the states mentioned above are localized at the two edges. Other states are extended in the x direction which can be considered as belonging to the bulk bands. In the course of the cycle we can not avoid exciting bulk electrons(holes) in the conduction(valence) band, which means the adiabatic cycle of the 1D system can not complete. We also recalculate the position-energy relationship in the FIG. 2 (c) in Ref \cite{Qi2008}. It's easy to see from FIG. 1(d) the edge states are well localized at the edges and never cross the bulk insulator, reach the other edge as in FIG. 2 (c) in Ref\cite{Qi2008}. So the 1D picture of the electron pump is completely misleading and conclusion that the charge is always pumped is not valid. It's clear the failure of this 1D electron pump is due to the excitation of bulk electrons and holes. Some authors try to use the 1D electron pump to explain the QSHE and avoid exiting the bulk electrons and holes at the same time. However,this attempt also failed. \\

\begin{figure}
  \includegraphics[width=3.7 in,clip=true]{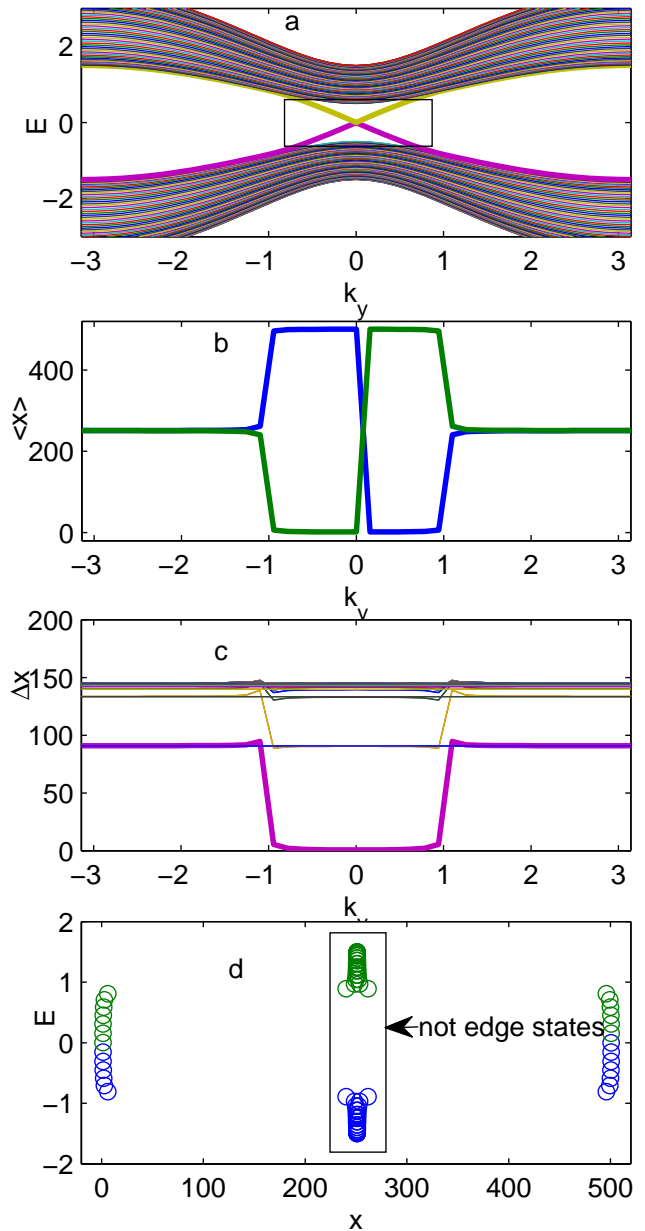}\\
  \caption{(a)The energy spectrum of the tight-binding model. (b)The average x position of the two branchs of eigenstates which are displayed by thick lines in (a). (c) the x position uncertainty $\Delta x=\sqrt{<(x-<x>)^2>}$ of all the eigenstates in (a). (d) the edge state position versus energy.}\label{f1}
\end{figure}

\subsection{A not so obvious mistake}
In ref\cite{Fu2006} a $Z_2$ spin pump is proposed to account the QSHE in 2D TIs. This pump is a cycle of a 1D system. So in a isolated system bulk electrons or holes will be excited in a cycle, which, according to the author, is because there is no ``space'' to put the excitations at the ends. To avoid this awkward situation several sites are added to the end of the system. By study this tight-binding model they come to the conclusion that ``For an isolated system, a single closed cycle of the pump changes the expectation value of the spin at each end even when spin-orbit interactions violate the conservation of spin. A second cycle, however, returns the system to its original state.''.\\
In their argument the system returning to the origin state after second cycle depends on the fact the four fold degeneracy of the first excited state at the end of the system when $t=0,T...$ will be split by the electron-electron interactions as shown in FIG. 1.(d) in Ref\cite{Fu2006}. There are plenty of examples that those kinds of degeneracies aren't all split by interaction. Since we talk about the end states of 1D system, the excited states of atoms or molecules provide good analogies. For example, we consider the first excited state of He atom. If the electron-electron interaction is neglected, the $(1s)(2s)$ states are four fold degenerate similar to the excited state at the end. However the strong Coulomb interactions don't split the degeneracy completely. The triplet states still degenerate.\\
More seriously, if the four fold degeneracy were split completely the first part of the conclusion will collapse. At $t=0,T...$ the Hamiltonian is time-reversal symmetric and if a state is eigenstate of the Hamiltonian the time reversed state will be a eigenstate with same energy. For a nondegenerate state, this means the state is time reversal symmetric. A time reversal symmetric state have zero average spins. So if the first excited state is nondegenerate a single closed cycle of the pump won't change the expectation value of the spin at the end as they claimed in the first part of the conclusion. \\
Here we face a dilemma that which part of the conclusion we should believe. If we accept that after one cycle there are nonzero average spins at the ends, the time reversed state must have same energy and opposite average spins. Clearly this state and the time reversed state are linearly independent. So there must be at least two fold degeneracy. The time reversed state can be arrived by a reverse cycle $0\rightarrow-T$. Since the pumping cycles we discussed are all reversible. Only when the state and the time reversed state are identical can the system return to the original ground state after the second cycle. However, as long as the average spin is nonzero those two states can't be identical. In this condition there will be a level crossing at $t=T$ and the system will go upward to another exited state instead of return to the ground state in the second cycle. All this can be can be clearly understood within the single electron picture as shown in FIG. 2. So we conclude as long as a single closed cycle can change the expectation value of the spin at each end the pump can't be a $Z_2$ type, that is it won't return to its original state after second cycle. \\
\begin{figure}
  \includegraphics[width=3.7 in,clip=true]{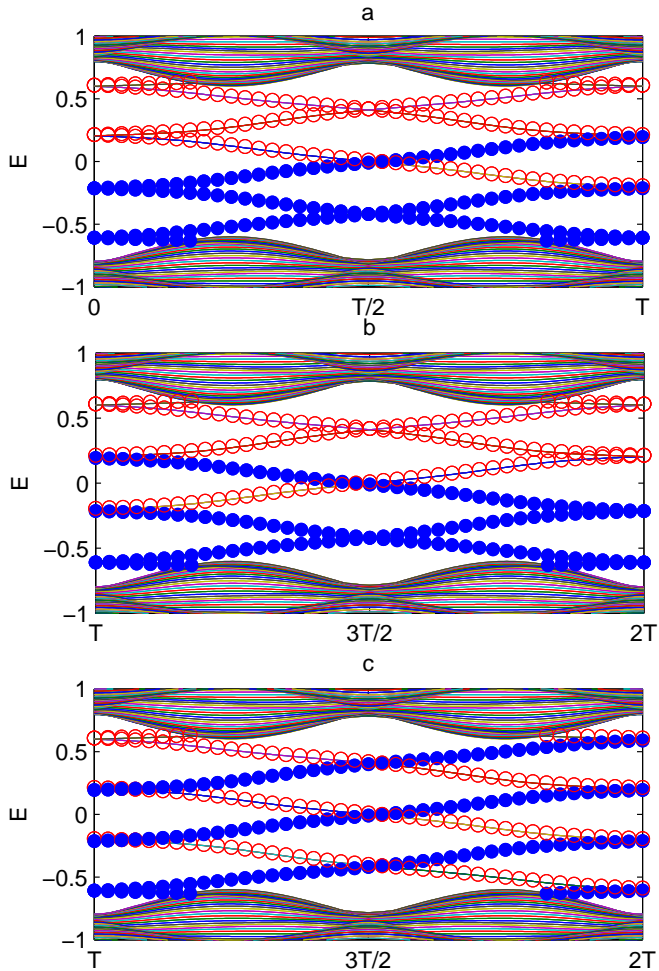}\\
  \caption{(a)One cycle of the 1D system begin with ground state. (b)One cycle of the 1D system end with ground state. If the state at $t=T$ in (a) is identical with the state at $t=T$ in (b) the system will come back to the ground state after two cycles. However, the two states are orthogonal at the single-electron level and have opposite average spins. (c)The second cycle follows the first cycle in (a) at the single-electron level. More spins are pumped after the second cycle.}\label{f2}
\end{figure}

If the second part of the conclusion is true, a single cycle can't pump spins. There won't be QSHE at all. For this artificial 1D system there is no experiment to test which part of the conclusion is true. Never mind, we now argue  we don't need these cycles of 1D system to explain QSHE in 2D TIs.
\section{Electron pumps in 2D system}
All the electron pumping mechanism in 2D insulators can be traced back to Laughlin¡¯s argument for integer quantum Hall effect\cite{Laughlin1981}.  In one sense the Laughlin¡¯s electron pump is much smaller comparing to the electron pumps for 1D systems in section I. For example, in the configuration in Ref\cite{Wu2011a} when magnetic flux threading the cylinder change by one flux quantum $k_2$ shifts by $2\pi/N$. In this process the Hamiltonian returns to its original form and the 2D system finish a cycle. Comparing to the cycle for 1D system where in one cycle $k_2$ change by $2\pi$, this cycle is really much smaller. There are many advantages in using this smaller electron pump. First, because we consider a 2D system there will be many edge states. Those state can be either occupied(bellow the Fermi level) or unoccupied(above the Fermi level). In one cycle the $k_2$ move only a little step. So we don't have to worry about exciting bulk electrons and holes as in section I A as long as the Fermi level is far away from the bulk bands . Nor do we need to add extra sites at the edge of the TIs to accommodate the pumped electrons as in section I B.\\
 In the clean limit when $k_2$ can be defined there are two equivalent ways to represent this small pump as illustrated in FIG. 3. The small pump can still be defined when $k_2$ can'e be defined. For example, when Translation symmetry is destroyed by the weak random time reversal symmetric potential or the 2D TIs form a Corbino disk instead of a cylinder, $k_2$ is not a good quantum number. We see from FIG. 3.(c) and (d) the topology of the edge states evolution is identical  to the clean limit in a cylinder. It's easy to see the small cycles have the same time reversal symmetry as the cycle of 1D systems because the Hamiltonians with opposite threading magnetic flux are time reversal related and the Hamiltonians are same when the magnetic flux difference is integer flux quantum. So, in those cases, the electron pump can still be considered as two time reversal related Chern pumps of Chern insulators.\\
 \begin{figure}
  \includegraphics[width=3.7 in,clip=true]{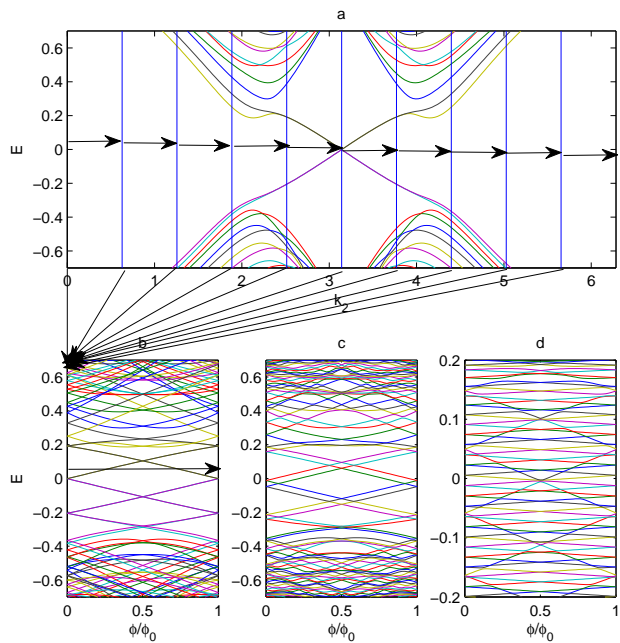}\\
  \caption{(a)An illustration of small pump. There are $N=10$ lattices along $\mathbf{a_2}$ direction in ref(). The vertical lines represent the $k_2s$ allowed by the period boundary condition. In one cycle the allowed $k_2$ move to $k_2+2\pi/N$. (b)Another illustration of the pump. All the eigenstates are moved to $\phi=0$. (c) A cycle when time-reversal random potential added to the system. There are double lines at the edge state area because the random potential destroy the inversion symmetry, the edge states at two edges no longer have same energy. (d)A cycle in a Corbino disk with inner diameter 3 and external diameter 18 in unit of lattice constant. It's easy to discriminate the inner and outer edge states.}\label{f3}
\end{figure}

 The small pumps have been used to explain the QSHE. For example, in the pioneering work of TI\cite{Kane2005b} its claimed that after a cycle ``However, in the QSH state a particle-hole excitation is produced at the Fermi energy $E_{f}$'' and ``spin accumulates at the edge''. So the problem left is whether the pump is a $Z_2$ type or Chern type, that is, after a second cycle the system return to its origin state or go to another excited state. They seem to agree with the $Z_2$ pump because ``if a second flux is added, then there will be T invariant interactions which do connect the state with the zero flux state. This suggests that the state with one flux added is distinguished by a $Z_2$ `T polarization'.'' However,  the same arguement in section I B can be applied here: if the first cycle do pump spin the system won't come back to its original state after the second cycle. So the small pump is certainly not a $Z_2$ type.\\
  Whether the pump is Chern type depends on the single electron picture works or not. The existing experiment result shows the single electron picture works pretty well\cite{Konig,roth09,Brune}. In Ref\cite{roth09} the transport property of quantum spin Hall state are calculated by using the Landauer-B\"{u}ttiker equation without invoking any electron-electron interaction. The calculation results are in good agreement of experiment results. In Ref\cite{Brune} the key of the experiment is to produce a the difference in chemical potential between the two spin states at the quantum spin Hall region. The different chemical potentials for different spin states is only meaningful when the single-electron picture works. Guaranteed the electron-electron interactions will cause some relaxation in the pumping process once the system is in excited state, as long as the single-particle picture is not totally destroyed the electron pumps should still be classified as Chern type because they continuously pumping spin to the edges.\\
\section{The relationship between $Z_2$ and Chern number}
As a beginner in the field I am baffled by the K theory and the mathematical jargons in Ref \cite{Fu2006}. I believe there are many researchers share the same experiences with me. Here we give a constructive definition of the $Z_2$ and establish the relation between $Z_2$ and Chern number. My version of the story assume little mathematical prerequisites and can be easily understood by the graduate students.\\
 For simplicity we consider the case only two bands are occupied. Because total Chern number of the vector bundle is zero, we can find two wave functions defined continuously on the brillouin zone to span the the bundle. We try to construct two continuous wave functions satisfy the time reversal constraint (A1) in Ref\cite{Fu2006}. The two continuously defined wave functions are trivial and have zero Chern number. So their 1D Wannier centers are continuous functions of $t$ and return to their original values when $t$ varies a cycle, which is illustrated in FIG. 4. To achieve our goal it's sufficient to find two wave functions continuously defined on the half of the brillouin zone and satisfy the time reversal constraint at the borders of the half of the brillouin zone with $t=-\pi$ and $t=0$. \\
  \begin{figure}
  \includegraphics[width=3.7 in,clip=true]{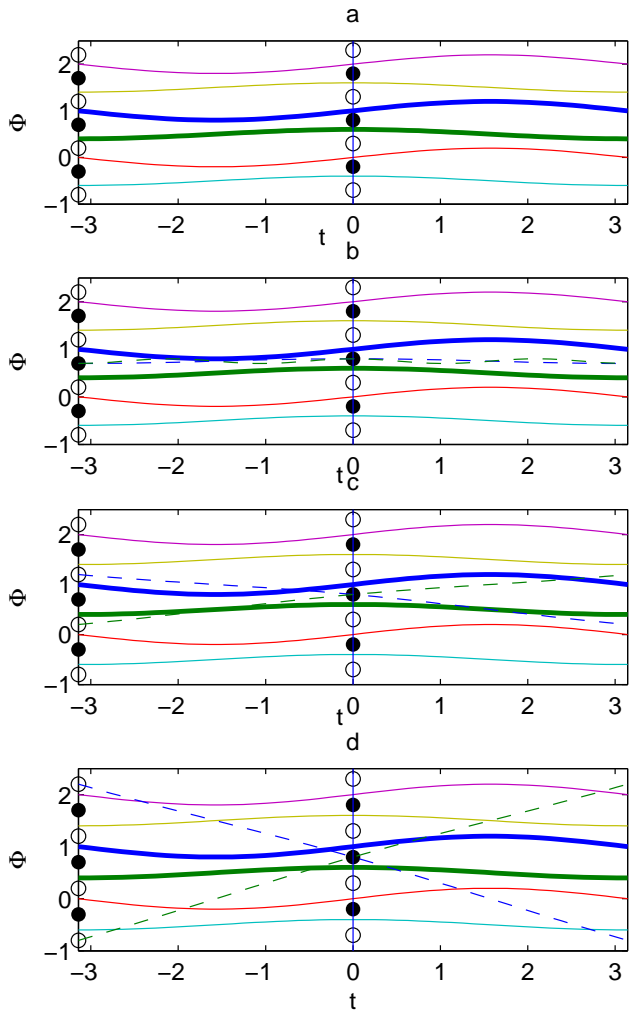}\\
  \caption{(a)The Wannier centers of the two continuous functions span the vector bundle.(b)The trivial insulator case.(c)(d)The TI case. }\label{f4}
\end{figure}
 Now we check whether it is possible. If the constraint is satisfied at $t=-\pi$ and $t=0$, the Wannier centers of the two functions will be identical there. Under general $U(2)$ transformation, the 1D charge polarization, which is the sum of the Wannier centers, can only be changed by integers. If we want to construct time reversal related functions from the trivial functions by $U(2)$ transformation at the edges, the centers of the T related functions can only be located at the points marked by dots and circles in FIG. 4.(a). However, as proven in Ref\cite{Fu2006}, the centers of the T related functions($P^I$ or $P^{II}$ in Ref\cite{Fu2006}) are defined modulo an integer. Then the $P^I$ can be represented by either the dots or the circles. To construct continuous a $U(2)$ transformation on the half brillouin zone, the two $U(2)$ transformations at $t=-\pi$ and $t=0$ must be continuously connected. This is only possible when those two $U(2)$ transformations belong to the same homotopy class. In our case it means the two $U(2)$ transformations should shift the 1D charge polarization by same integer. If the $P^I$ at the two edges are of the same type, e.g. they are both marked by dots, the two $U(2)$ transformations constructing the desired wave functions at the two edges can be continuously connected on half of the brillouin zone. In this way we construct the desired wave functions on the half zone.
 Then we apply the time reversal operator to the wave functions to obtain the definition of the wave functions on the other half of the brillouin zone. We find two trivial wave functions satisfy the time reversal constraint and the insulator is trivial.\\
 If $P^Is$ at $t=-\pi$ and $t=0$ are not of same type, e.g. they are represented by dots at $t=0$ and circles at $t=-\pi$ as in FIG. 4.(c) and (d), to reach the T related pairs at the two edges the total charge polarization must shift different integers and the difference is odd. So no $U(2)$ transformations of the same homotopy class at the edges can be defined to construct the desired wave functions. In this case $Z_2=1$ and the insulator is topological.\\
 Now we discuss to what extent can the wave functions satisfy the T restraint be defined in the TI. It's easy to see $U(2)$ transformations of the same homotopy can only move the Wanier centers to the location of $P^I$ as illustrate in FIG. 4. (c) and (d). The transformed wave functions at $t=-\pi$ are still related by the time reversal operator as in (A1) in Ref\cite{Fu2006} except with an extra $U(1)$ phase factor. The phase variation when $k$ moves a cycle is $C2\pi$, $C$ is an odd integer. We apply the time reversal operator to the transformed wave functions to get the definition of wave functions on the other half of the Brillouin zone. Then we get wave functions satisfy the T restraint. However, as shown in FIG. 4. (c)and(d), those two wave functions aren't continuous. Their 1D Wannier centers change by $\pm C$ when $t$ varies from $-\pi$ to $\pi$. The Chern numbers of the two wave functions are $\pm C$. Thus, we conclude the only wave functions with opposite odd Chern numbers can be constructed in TIs if the time reversal restraint is enforced on the two functions. The conclusion can be easily generalized to more than two bands. Similar conclusions have been mentioned by several authors\cite{Soluyanov2011,Roy}. However, their discussion are all limited to specific models and the equivalence of the $Z_2$ and the parity of the Chern number hasn't been proven.\\
From above discussion it's easy to see there are arbitrariness in constructing the time reversal wave functions. In our previous work we proposed a way to construct the wave functions uniquely for a given edge. With our method we can not only determine the topology of the edge states but also get some geometrical information of the edge states. In our view the 2D TI can be considered as two Chern insulators with odd Chern number even when the spin is not conserve. The two groups of T related edge state corresponding to the two Chern insulators. Instead of return to the original state after two cycles the Chern insulators can continuously pump electrons to opposite direction , thus produce the QSHE. This is the biggest difference between our picture and $Z_2$ picture. Since we have shown the $Z_2$ pump theory is self-contradictory and the single-electron picture works well in explaining the experimental result. We believe it's very possible that our theory is the true answer to the problem.\\
Another popular theory is to describe the 2D TIs by the spin-Chern number\cite{sheng,Prodan}. The spin-Chern number have been proposed, dismissed and redefined. We will discuss the redefined final version. Here only the clean limit when Brillouin zone can be defined is considered. In Ref \cite{Prodan} the eigenstates of operator $P(\mathbf{k})\hat{s_{z}}P(\mathbf{k})$ are used to calculated the spin Chern number. The Chern number of eigenstates corresponding to the positive and negative eigenvalues of the $P(\mathbf{k})\hat{s_{z}}P(\mathbf{k})$ are calculated to define the spin Chern number. It's a specific way to decompose the occupied bands with the-time reversal constraint. If $|u_I(\mathbf{k})>$ is the eigenstate of $P(\mathbf{k})\hat{s_{z}}P(\mathbf{k})$ with eigenvalue $\lambda$, $\Theta|u_I(\mathbf{k})>$ will be eigenstate of $P(-\mathbf{k})\hat{s_{z}}P(-\mathbf{k})$ with eigenvalue $-\lambda$. So the states with positive and negative eigenvalues satisfy the time reversal restraint. The problem of spin-Chern number is that there is no unique way to choose the spins. When $\hat{s_z}$ isn't conserve there is no good reason we must choose $P(\mathbf{k})\hat{s_{z}}P(\mathbf{k})$ to decompose the occupied band. In fact $P(\mathbf{k})(\hat{s_{x}}+0.1*\hat{s_{z}})P(\mathbf{k})$ and $P(\mathbf{k})(\hat{s_{y}}+0.1*\hat{s_{z}})P(\mathbf{k})$ will do the same work as shown in FIG. 5. This arbitrariness indicated the spin-Chern number is just a equivalent expression of $Z_2$, it can't provide more information than $Z_2$. In comparison, our method of decomposition is unique for a given edge and provide some geometrical information of the edge states.\\
  \begin{figure}
  \includegraphics[width=3.7 in,clip=true]{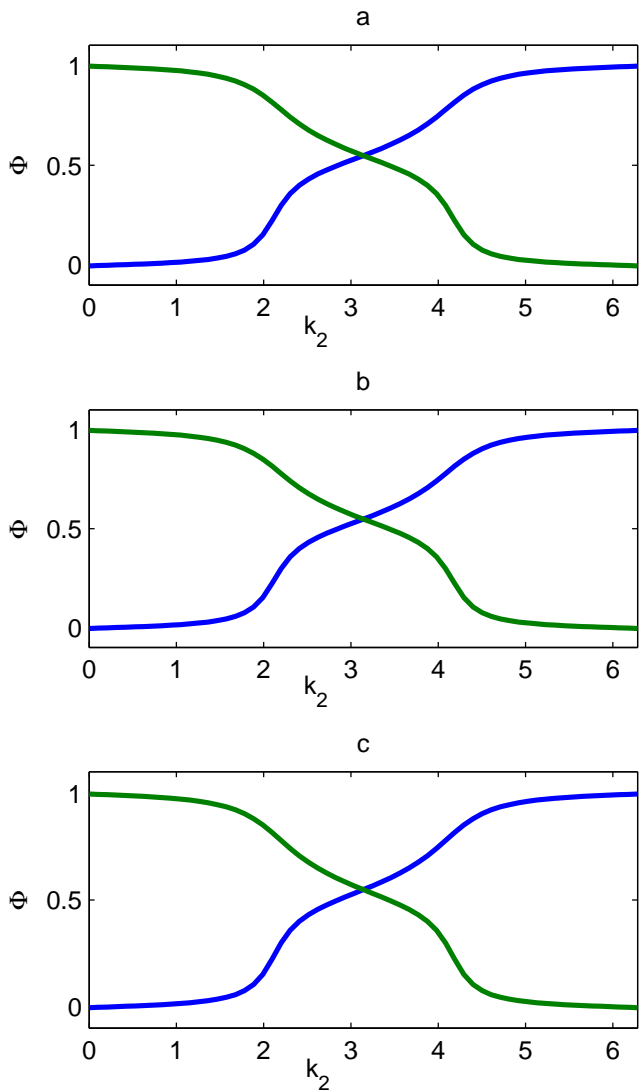}\\
  \caption{(a)(b)(c)The Wannier centers of the decomposed bands as eigenstate of $P(\mathbf{k})\hat{s_{z}}P(\mathbf{k})$ $P(\mathbf{k})(\hat{s_{x}}+0.1*\hat{s_{z}})P(\mathbf{k})$ and $P(\mathbf{k})(\hat{s_{y}}+0.1*\hat{s_{z}})P(\mathbf{k})$ respectively. They all decompose the bands to two bands with Chern number $\pm1$. We use the Kane and Mele model with parameters $\lambda_{SO}=0.06$, $\lambda_{r}=0.05$ and $\lambda_{v}=0.1$  }\label{f5}
\end{figure}
One of the referees point out ``The adaptation of Laughlin's argument to the $Z_2$ TI has also been explicitly discussed'' in Ref\cite{essin}. Since the total Chern number of the occupied bands of TI is zero, the real $Z_2$ insulator can't pump charges in any pumping process. The claim ``While this definition precisely reproduce the supercell definition above, it is slightly different from the pumping in the IQHE since, in some cases, the second stage requires changing the system's Hamiltonian and not just the boundary'' in Ref\cite{essin} seems a little inaccurate. The truth is for this pump to work in any case there must be big change of Hamiltonian of the TI because the total Chern number of the Hamiltonian must change and the time reversal symmetry of the system must be broken. We don't believe this can be deemed as slightly different from a pumping in IQHE. So this pump can never be realized in real 2D TIs.
\section{Suggestions for the experimental physicist }
We do not just provide an equivalent or refined version of the $Z_2$ theory. Our aim is overturn the $Z_2$ theory of QSHE completely. For this purpose we make concrete experimental predictions in our previous work. We predict the QSHE can be observed in isolated devise e.g. a Corbino disk. Our suggestion is an genuine test for QSHE because we predict there are spin current perpendicular to electrical field created the changing magnetic flux and spins will accumulate at the two edges in isolate devise. In comparison the recent experiment work using an external device (metal spin hall device)  to create the Fermi level difference of the spins at the edge, so the spin accumulation at edge is not due to the intrinsic QSHE in the TI.\\
Also we provide a new way to looked the QSHE. In a recent paper\cite{Zhou} QSHE is described as ``The QSHE can be viewed as being the net result of two distinct quantum Hall states, or rather edge channels ¡ª one of only spin-up electrons traveling in one
direction, and the other of only spin-down electrons traveling along the same edge path but in exactly the opposite direction.'' In this description it's clear the author believe the spins are transported through the edge channel and it's a common belief. On the contrary in our suggestion the spin current is in the bulk and the edges serve as reservoirs to accommodate the pumped spins.\\
If our previous suggestion isn't novel enough we now go a step further. All the calculations of edge states up to now show the gapless edge state in TIs connect the occupied valence band and the unoccupied conduction band. If the single-election picture works when we continuously increase the magnetic flux the two time reversal related Chern pump bound to pump electrons from the bulk valence band to the conduction band. Even if the electron-electron interaction cause some relaxation, as long as the pumping rate exceed the relaxation rate the bulk electron hole pairs can still be created. Once the electrons or holes enter into the bulk bands the process cease to be adiabatic because they will quickly become delocalized. Then the bulk electrons and holes won't relax through the edge channels. The bulk electrons and holes will recombine and emit light. We call this process topological luminescence. If our theory works this phenomenon can be observed in a Corbino disk as illustrated in FIG. 3.(d). However, the experiment requirement isn't so demanding. We can also observe this phenomenon in a disc or square with the magnetic flux pass through the bulk. We show the evolution of the edge states of a disc and a square when the magnetic flux varies in FIG. 6. The magnetic flux need not be continuously increasing. Once the bulk electron or hole is create, it won't be pumped back by a reversed cycle. So an alternating  magnetic field may also continuously generate bulk electrons and holes.
   \begin{figure}
  \includegraphics[width=3.7 in,clip=true]{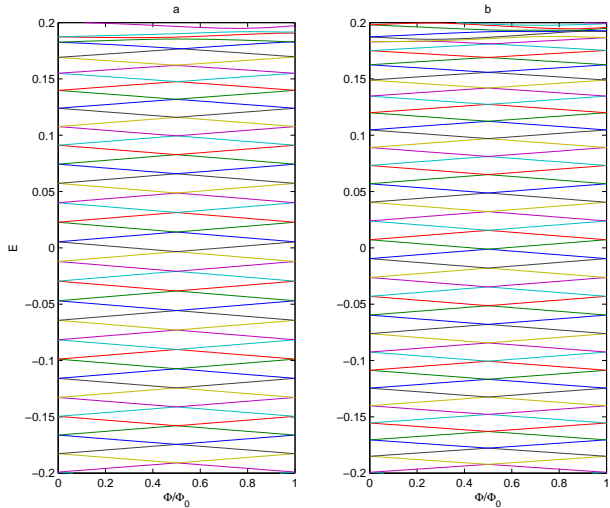}\\
  \caption{(a)Edge states evolution with magnetic flux in a disc of radius 18 in unit of lattice constant. The magnetic flux are uniformly distributed within a circle of radius 5. (b)Edge states evolution with magnetic flux in a $35\times35$ square. The magnetic flux is same as in (a). }\label{f6}
\end{figure}


\begin{thebibliography}{10}

\bibitem{Wu2011a}
Y.D. Wu, arXiv:1208.4096(unpublished).

\bibitem{Kane2005a}
C.L. Kane and E.J. Mele 
Phys. Rev. Lett. \textbf{95}, 226801 (2005).

\bibitem{Bernevig2006}
 B.A. Bernevig, T.L. Hughes and S.C. Zhang, 
Science \textbf{314}, 1757-1761 (2006).

\bibitem{Kane2005b}
C.L. Kane and E.J. Mele, 
Phys. Rev. Lett. \textbf{95}, 146802 (2005).

\bibitem{Hasan2010}
M.Z. Hasan and C.L. Kane, 
Rev. Mod. Phys. \textbf{82}, 3045 (2010).


\bibitem{Qi2008} X.L. Qi, T.L. Hughes, and S.C. Zhang, Phys. Rev. B {\bf 78}, 195424 (2008).

\bibitem{Fu2006}
L. Fu and C.L. Kane,Phys. Rev. B. \textbf{74}, 195312 (2006).


\bibitem{Laughlin1981}
R.B. Laughlin, Phys. Rev. B \textbf{23}, 5632-5633 (1981).


\bibitem{Konig} M. K\"onig,, S. Wiedmann, C. Br\"{u}ne, A. Roth, H. Buhmann, L. W. Molenkamp, X. L. Qi and S. C. Zhang, 2007, Science {\bf 318}, 766(2007).

\bibitem{roth09}
A. Roth, C. Br\"une, H. Buhmann, L. W. Molenkamp, J. Maciejko, X. L. Qi, and S. C. Zhang, 2009,
Science {\bf 325}, 294(2009).


\bibitem{Brune}
C. Br\"une,, \emph{et~al.} Nat. Phys. \textbf{8}, 485 (2012).

\bibitem{Soluyanov2011}
A.A. Soluyanov and D.Vanderbilt, \textit{Phys. Rev. B}. \textbf{83}, 035108 (2011).

\bibitem{Roy}R. Roy,  Phys. Rev. B {\bf 79}, 195321 (2009).

\bibitem{sheng} L. Sheng, D.N. Sheng, C.S. Ting and F.D.M. Haldane, Phys. Rev. Lett. {\bf 95}, 136602 (2005).



\bibitem{Prodan}E Prodan,  Phys. Rev. B {\bf 80}, 125327(2009).

\bibitem{essin} A. M. Essin,, J. E. Moore and D. Vanderbilt, Phys. Rev. B {\bf 76} 165307.

 \bibitem{Zhou}
 Y. Zhou and F.C. Zhang Nat. Phys. \textbf{8}, 448 (2012). 








\end{thebibliography}
\end{document}